\documentclass{PoS}
\usepackage{graphicx}

\def\MeV {\mathop{\hbox{MeV}}}

\def\Re {\mathop{\hbox{Re}}}

\def\DU  {\mathop{{\cal D}\hbox{U}}}

\def\dd  {\mbox{d}}
\newcommand\detn[1]{\mbox{det}_{#1}}

\newcommand{\beq}{\begin{equation}}
\newcommand{\eeq}{\end{equation}}
\newcommand{\beqa}{\begin{eqnarray}}
\newcommand{\eeqa}{\end{eqnarray}}

\title{Finite Density Simulation with Canonical Ensemble}

\ShortTitle{Finite Density Simulations with Canonical Ensemble}
\author{$\chi$QCD Collaboration}
\author{\speaker{Anyi Li}\\
        Department of Physics and Astronomy, University of
Kentucky, Lexington KY 40506, USA\\
        E-mail: \email{anyili@pa.uky.edu}}

\author{Xiangfei Meng\\
        Department of Physics and Astronomy, University of
Kentucky, Lexington KY 40506, USA\\
Department of Physics, Nankai University, Tianjin 300071, China
\\Email: \email{mengxf@mail.nankai.edu.cn}
        }

\author{Andrei Alexandru\\
        Physics Department, The George Washington University, Washington, DC 20052, USA\\
        E-mail: \email{aalexan@gwu.edu}
        }
\author{Keh-Fei Liu\\
        Department of Physics and Astronomy, University of
Kentucky, Lexington KY 40506, USA \\
E-mail: \email{liu@pa.uky.edu}
        }
\abstract{ QCD at non-zero baryon density is expected to have a
critical point where the zero-density cross-over turns into a
first order phase transition. To identify this point we scan the
density-temperature space using a canonical ensemble method. For a
given temperature, we plot the chemical potential as a function of
density looking for an ``S-shape'' as a signal for a first order
transition. We carried out simulations using Wilson fermions with
$m_\pi \approx 1\mbox{GeV}$ on $6^3\times 4$ lattices. As a
benchmark, we ran four flavors simulations where we observe a
clear signal. In the two flavors case we do not see any signal for
temperatures as low as $0.83 T_c$. Preliminary results for the
three flavor case are also presented. }

\FullConference{The XXVI International Symposium on Lattice Field Theory\\
         July 14-19 2008\\
         Williamsburg, Virginia, USA}

\begin{document}

\section{Introduction}

In recent years, full QCD simulations have become feasible due to
development of new algorithms and increasing computational power.
Lattice simulations using dynamical fermions can now be performed
at finite temperature and zero baryon density. However,
simulations at non-zero baryon density remain a challenge for lattice QCD
due to the complex nature of the fermionic determinant where the
conventional Monte Carlo methods fail. The standard solution of
splitting the action into the real and positive part and an extra
phase fails due to sign and overlap problems.
To address the overlap problem, a method based on the canonical
partition function has been proposed~\cite{kfl05}. While expensive --
every update involves the evaluation of the fermionic determinant --
finite baryon density simulations based on this method proved feasible~\cite{afhl05}
and a program was
outlined to scan the QCD phase diagram to look for the critical point~\cite{lal06,lal07}.

In this paper, we present results based on simulations on
$6^3\times 4$ lattices with Wilson fermions. We plot the chemical potential
as a function of baryon density and we clearly observe the ``S-shape''
structure in the $N_f = 4$ case, indicating a first order phase
transition~\cite{fk06}. We do not see such a structure in the $N_f
= 2$ case down to $0.83T_c$. We will also present preliminary
results for $N_f = 3$.

\section{Algorithm}
The simplest way to show how to build the canonical ensemble in
Lattice QCD is to start from the fugacity expansion,
\beq
Z(V,T,\mu) = \sum_{k} Z_C(V, T, n) e^{\mu k/T}, \label{fugacity}
\eeq
where $k$ is the net number of quarks (number of quarks minus
the number of anti-quarks) and $Z_C$ is the canonical partition
function. Using the fugacity expansion, it is easy to see that we
can write the canonical partition function as a Fourier transform
of the grand canonical partition function,
\beq
Z_C(V, T, k) =
\frac{1}{2\pi} \int_0^{2\pi} \mbox{d}\phi \,e^{-i k \phi} Z(V, T,
\mu)|_{\mu=i\phi T}.
\eeq

As an illustration, we will consider the case of two degenerate
flavors. After integrating out the fermionic part, we get a simple
expression \beq Z_C(V, T, k) = \int \DU e^{-S_g(U)} \detn{k}
M^2(U)\label{eq:canonical}, \eeq where \beq \detn{k} M^2(U) \equiv
\frac{1}{2\pi}\int_0^{2\pi} \dd\phi\,e^{-i k \phi} \det M(m,
\mu;U)^2|_{\mu=i\phi T} , \eeq is the  projected determinant with
the fixed net quark number $k$. $\detn{k}M^2(U)$ is a real number
due to the charge conjugation symmetry of the canonical partition
function. However, it is not necessarily positive. In our
simulations, we use $|\Re \detn{k}M^2(U)|$ and fold the phase
factor in the observables.

Exact determinant calculation of fermion matrix is very
demanding even on $6^3\times4$ lattices. An alternative is
to use a noisy estimator~\cite{all07,jhl03} but this is quite cumbersome.
For simplicity sake, we used exact evaluation of the determinant in this study.
Another technical problem has to do with the Fourier transform; our original
approach was to use an approximation where we replaced the continuous
definition with a discrete one, i.e.:
\begin{equation}
\detn{k} M^2(U) \approx \frac{1}{N} \sum_{j=0}^{N-1} e^{-i k
\phi_j} \det M(U_{\phi_j})^2,~~~~~\phi_j=\frac{2\pi
j}{N}.\label{eq:fourier}
\end{equation}
It was shown that the errors introduced by this approximation are small~\cite{lal07}
at least for small quark numbers. There are two problems with this approach:
the computation time increases linearly with the net quark number and, for large
enough densities, the Fourier components become too small to be evaluated using
double precision floating point numbers. To address these issues, in this study we
used a different approximation method: the winding number expansion~\cite{mlal08}.
This method is both faster and more accurate than the original method.

As mentioned above, we used the chemical potential to probe the
phase transition. In the canonical approach the chemical potential
is measured as the increase in free energy when we introduce one
more baryon in the system, i.e.:
\begin{equation}
\label{baryon chemical potential} \left<\mu\right>_{n_B}   =
\frac{F(n_B+1)-F(n_B)}{(n_B+1)-n_B} = -\frac{1}{\beta}\ln
\frac{Z_C(3n_B+3)}{Z_C(3n_B)} = -\frac{1}{\beta}\ln \frac{\left<
\gamma(U)\right>_o}{\left< \alpha(U)\right>_o}
\end{equation}
where
\begin{eqnarray}
\alpha(U) &=& \frac{\Re\detn{3n_B} M^2(U)}{\left|\Re \detn{3n_B}
M^2(U)\right|} \quad {\rm and} \\
\gamma(U)&=&  \frac{{\rm Re\,}\detn{3n_B+3} M^2(U)} {\left|{\rm
Re\,}\detn{3n_B} M^2(U)\right|}.
\end{eqnarray}
$\alpha(U)$ is the phase and $\left<\right>_o$ stands for the
average over the ensemble generated with measure $\left|\Re
\detn{3n_b} M^2(U)\right|$. If the configurations in the ensemble
have equal probability for both positive and negative $\alpha(U)$,
the denominator in the equation above becomes zero and we have a
``sign problem''. In our simulations the sign oscillations are
under control.

\section {QCD phase diagram}

\begin{figure}[hbt!]
\centering
\includegraphics[scale=0.55]{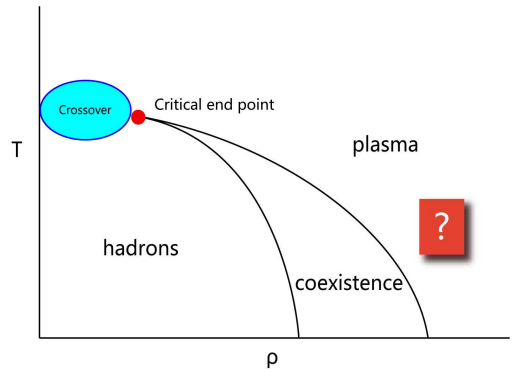}
\caption{Schematic phase diagram of QCD.\label{fig:1}}
\end{figure}

At zero baryon density, it has been known for quite some time that
QCD undergoes a transition from a confined phase to a deconfined
phase at a temperature $T_c\approx 170\MeV$. Lattice QCD suggests
that the transition is in fact a smooth crossover. This is
expected to turn into a first order phase transition as the baryon
density is increased. A schematic picture of the expected phase
diagram in the density-temperature plane (see Fig.~\ref{fig:1}) shows the crossover
ending with a critical point at some non-zero baryon density. The first order phase
transition is characterized by a coexistence region separating the hadronic phase
and the plasma phase.

To search for the phase boundaries of the coexistence region, we
scan the phase diagram by varying the density while keeping the
temperature fixed. The baryon chemical potential should exhibit an
``S-shape'' as one crosses the coexistence region~\cite{fk06}.

\begin{figure}[hbt!]
\centering
\includegraphics[scale=0.55]{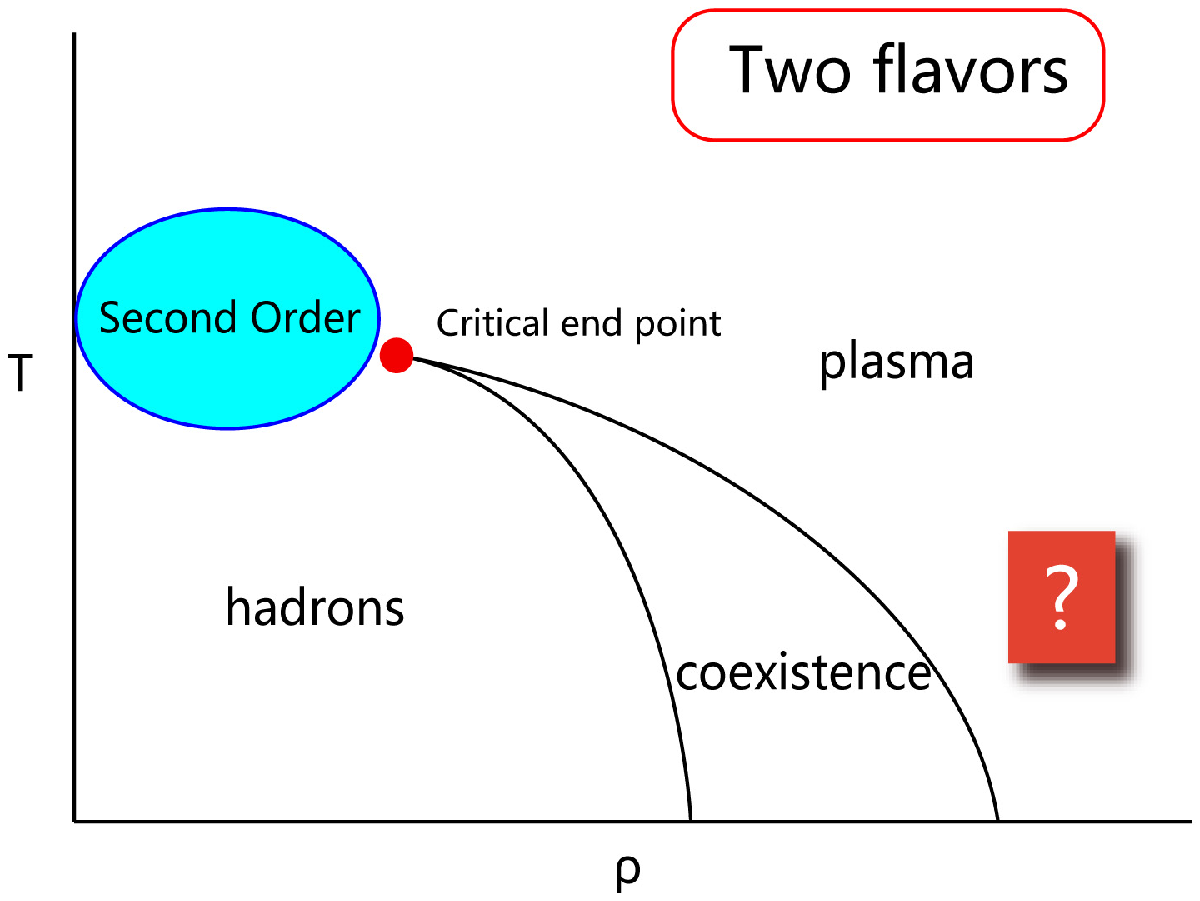}
\includegraphics[scale=0.55]{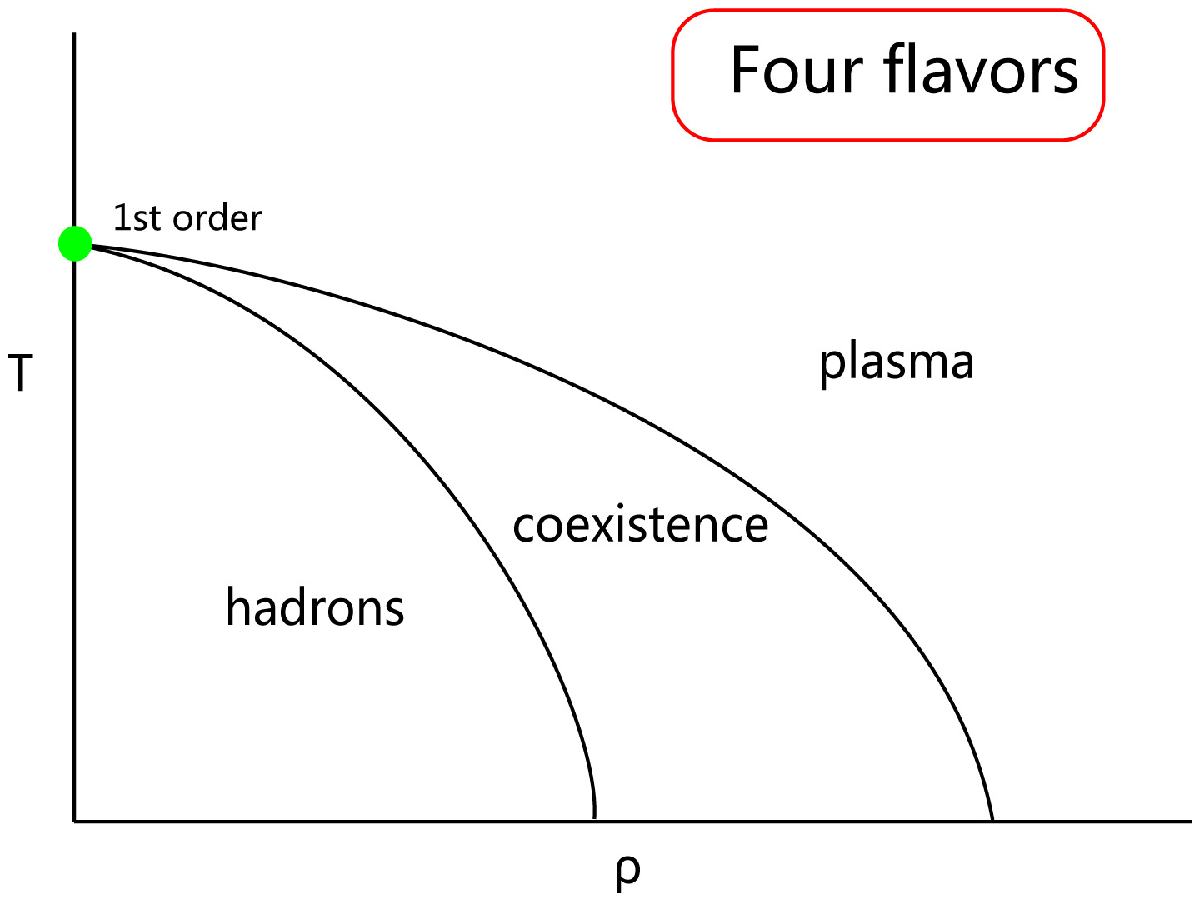}
\caption{Expected phase diagrams of two flavors (left) and four
flavors (right).\label{fig:2}}
\end{figure}

In our simulations we use quarks that are much heavier than the
physical quarks. Furthermore, for simplicity sake, we carry out
simulations where the quark masses are degenerate. We will present
results for simulations using 2, 3 and 4 degenerate flavors of
quarks. The expected phase diagrams for 2 and 4 flavor cases are
shown in Fig.~\ref{fig:2}. The two flavor case is expected to have
a diagram very similar to the full QCD one whereas it is known
that in the case of four flavors the first order phase transition
extends all the way to zero baryon density. We will use the four
flavor simulations as a benchmark to show that the methodology we
use can determine the boundaries of the coexistence region.

\section{Results}

\begin{figure}[hbt!]
\centering
\includegraphics[scale=0.45]{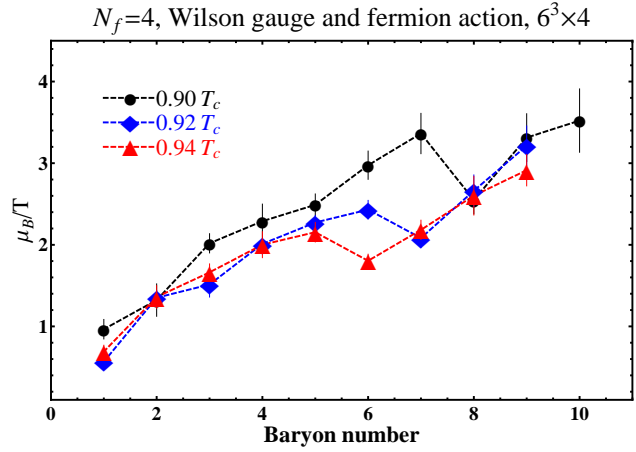}
\caption{Baryon chemical potential vs. baryon number at different
temperatures for $N_f=4$.\label{fig:3}}
\end{figure}

In Fig.~\ref{fig:3} we show the results for our four flavor
simulation. On the technical side, we note that we didn't have a
sign problem: even the simulations at the largest density where
the box size is $1.8{\rm fm}$ and the temperature is $0.90 T_c$,
the sign oscillations were moderate. The plots show a clear signal
for a first order phase transition when the temperatures are lower
than $T_c$.

To identify the boundaries of the coexistence region and the
critical value for the chemical potential we used the Maxwell
construction~\cite{fk06}. More precisely, we selected four points
in the ``S-shape"  region and we fit these points with a third
order polynomial. A better approach would be to use some
phenomenologically motivated functional form and try to fit a
larger region; we changed the fit function and we also extended
the fit region. For all reasonable fits, we found that the values
of the boundary points, $\rho_1$ and $\rho_2$, and the value of
the critical chemical potential, $\mu_c$, are fairly insensitive
to our choice of the fit function or fit region -- the simple
third order polynomial fit was sufficient.

\begin{figure}[hbt!]
\centering
\includegraphics[scale=0.45]{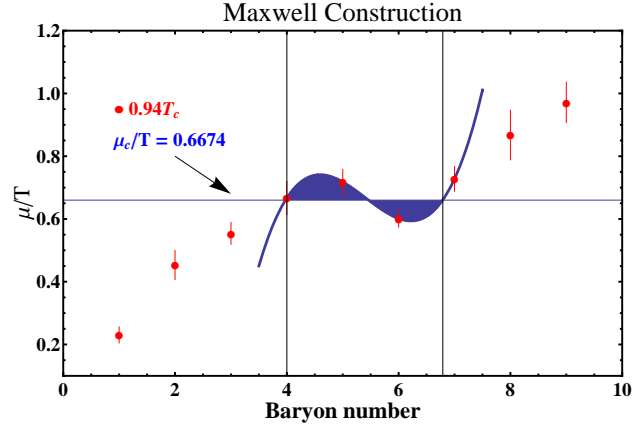}
\caption{Maxwell construction for $T=0.94T_c$ and $N_f=4$.}
\end{figure}

\begin{figure}[hbt!]
\centering
\includegraphics[scale=0.5]{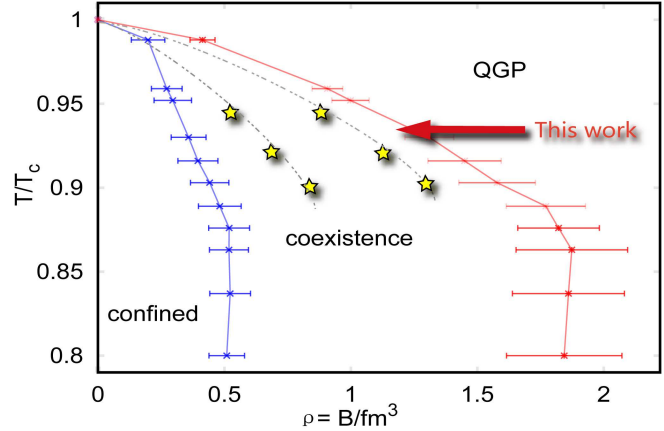}
\caption{Phase boundary for four flavors QCD: our work and the
results of a study using staggered fermions~\cite{fk06}.
(Dashed lines are plotted to guide the eye; they are not the result
of an extrapolation. The error bars are not determined yet.)\label{fig:5}}
\end{figure}

Once the $\rho_1$ and $\rho_2$ are determined for different
temperatures, they can be used to plot the boundaries of the
coexistence region. We can also find the critical point by
determining where the width of the coexistence region shrinks to
zero. In Fig.~\ref{fig:5} we compare our results to those from a
study using staggered fermions~\cite{fk06}. We see that our
coexistence region is much narrower. This could be due to our
heavier pion mass ($m_\pi \approx 1 \mbox{GeV}$ compare to $m_\pi
\approx 300 \mbox{MeV}$), or due to the fact that we use a
different fermion formulation.

The fact that the results of our four flavors simulations are consistent with the
expectation and with other lattice studies is encouraging. The only issue that
needs to be addressed is the discrepancy in the location of the boundaries.

\begin{figure}[hbp!]
\centering
\includegraphics[scale=0.45]{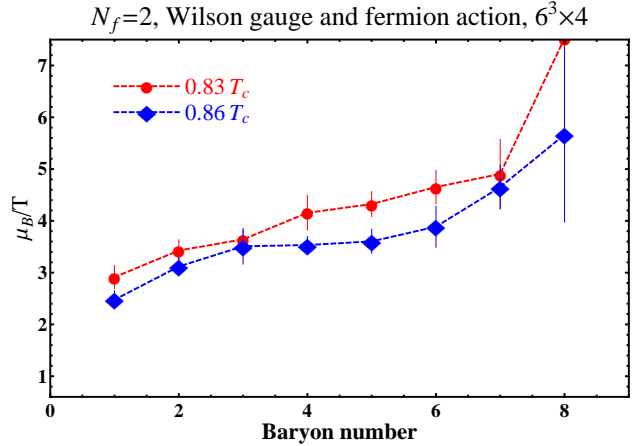}
\caption{Baryon chemical potential vs. baryon number for $N_f=2$.\label{fig:6}}
\end{figure}

In Fig.~\ref{fig:6} we show our results for $N_f=2$ simulations. These simulations
are more expensive than the four flavors simulations due to sign fluctuations. In the
two simulations that we ran at $T=0.86T_c$ and at $T=0.83T_c$ we do not see any signal
for a first order phase transition. There is at least one claim~\cite{Ejiri:2008xt} that the critical point
occurs at temperatures as low as $T=0.8T_c$. If this is indeed the case, we need
to run simulations at even lower temperatures in order to see the ``S-shape".

Three quark flavors are relevant for the structure of matter at
energies of the order of few hundred MeV: two light quarks $u$,
$d$ and one heavier quark $s$. Lattice simulations that come close
to approximate full QCD treat the lighter quark flavors as
degenerate and use one heavier quark flavor. Ideally, we would
like to carry out simulations close to the physical point.
However, this is not really practical especially for Wilson
fermions on lattices as coarse as we use in our study. In fact,
our quark masses are even heavier than the strange quark mass. We
decided to investigate the $N_f=3$ with the hope that the phase
diagram is, at least qualitatively, close to the full QCD phase
diagram.

\begin{figure}[thbp!]
\centering
\includegraphics[scale=0.45]{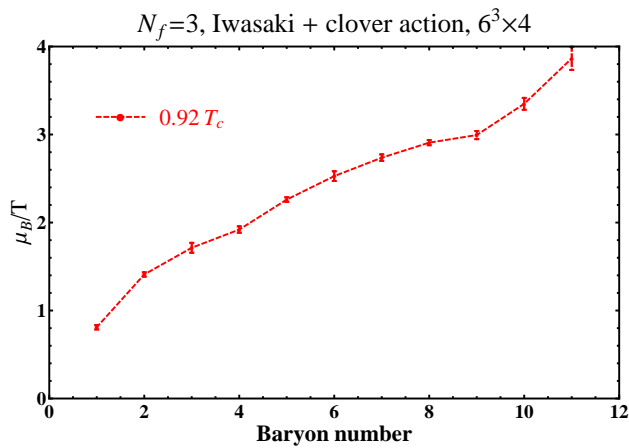}
\caption{Baryon chemical potential vs. baryon number at different
temperatures for four flavors. Simulations are done using Iwasaki
gauge action and clover fermion action.\label{fig:7}}
\end{figure}

In Fig.~\ref{fig:7} we show the results of our simulations for $N_f=3$.
For these simulations we used Iwasaki gauge action and clover fermions
in order to reduce lattice discretization errors. For $T=0.92 T_c$, we don't see any
signal for a first order phase transition.

\section{Summary}

We presented results for QCD simulations at non-zero baryon
density using the canonical ensemble approach. In the four flavor
case we see a clear signal for a first order phase transition --
this is consistent with phenomenological expectations and previous
lattice results. This proves that our method is sound. In the two
flavor case, we do not see any signal for temperatures as low as
$0.83 T_c$. We also presented results for three flavors where we
do not find any signal for a transition at $T=0.92 T_c$. We plan
to continue our investigations at lower temperatures and smaller
quark masses.

\end{document}